\documentclass[letterpaper,twocolumn,superscriptaddress,floatfix]{revtex4}
\usepackage[latin1]{inputenc}
\usepackage{bm}
\usepackage{multirow,amssymb,amsbsy,amsmath}
\usepackage{graphicx}
\usepackage{verbatim}
\makeatletter
\usepackage{pifont}
\makeatother

\begin{document}

\title{Experimental observation of optical precursors in optically pumped crystals}

\author{Zong-Quan Zhou}
\affiliation{Key Laboratory of Quantum Information, University of
Science and Technology of China, CAS, Hefei, 230026, China}
\author{Chuan-Feng Li$\footnote{email:cfli@ustc.edu.cn}$}
\affiliation{Key Laboratory of Quantum Information, University of
Science and Technology of China, CAS, Hefei, 230026, China}
\author{Guang-Can Guo}
\affiliation{Key Laboratory of Quantum Information, University of
Science and Technology of China, CAS, Hefei, 230026, China}

\date{\today}

\begin{abstract}
{We experimentally observed optical precursors in optically pumped crystals using polarization-based interference. By switching the user-programmable medium among the fast light, slow light and no-dispersion regimes, we observed an unchanged polarization states for the wavefronts. The robust polarization-encoded information carried by wavefronts suggests that precursors are the preferred carriers for both quantum and classical information in communication networks.}
\end{abstract}
\maketitle

Pulse propagation through a highly dispersive medium often leads to counterintuitive behavior. The motion of a pulse peak can be approximated by the group velocity ($\upsilon_{g}$), which is given by $\upsilon_{g}=c/(n+\omega dn/d\omega), $ where $c$ is the speed of light in a vacuum, $n$ is the refractive index of the material and $\omega$ is the frequency of the light. For the strong anomalous dispersion conditions where $dn/d\omega$ is negative, $\upsilon_{g}$ can be greater than $c$ and can even be negative \cite{Boyd09,fl1985,Wang00,Boyd03,Gauthier03,Dolling06,Boyd06}.
To resolve the apparent contradictions between fast light propagation and the theory of relativity, optical precursors were first introduced by Sommerfeld and Brillouin in 1914 \cite{Brillouin1960}. The theory states that the front edges of an ideal step-modulated pulse propagate at speed $c$ because of the finite response time of any physical material and that no components can overtake this wave front. Optical precursors are currently attracting renewed interest not only for fundamental reasons related to causality but also for possible applications in the generation of high peak-power optical pulses \cite{Segard,Jeong10,Du10}. Several experiments with different physical systems have confirmed that the speed of optical precursors (information) is equal to the velocity of light in a vacuum or in the background medium, independent of the group velocity \cite{Jeong06,Du08,Du09,Du11,Gauthier03,Gauthier05,Medium,Oishi,gisin04,sf}.

Due to the long optical coherence time and the large optical bandwidth, rare-earth-doped solids provide a highly controllable absorption profile, which has already found significant applications in quantum networks \cite{computation,eff10,AFC08,bw3,timebin}. The user-programmable absorption also provides a powerful platform for exploring the properties of optical precursors. We have previously proposed that polarization-based interference can be used to accurately determine the speed of optical precursors in rare-earth-doped solids \cite{scheme}.
In this Brief Report, we experimentally observed optical precursors using polarization-based interference in a user-programmable medium: optically pumped Nd$^{3+}$:YVO$_{4}$ crystal. The media in the two arms of the interferometer are independently programmed to be under very different dispersion conditions. We found that the polarization of the wavefronts is well-protected from the decoherence induced by the strong dispersions, which may find applications in both quantum and classical communication networks.

The experimental sample is a Nd$^{3+}$:YVO$_4$ crystal (doping level: 10 ppm). The length of the crystal is $L=$ 3 $mm$ along the a-axis. The $^{4}I_{9/2}\rightarrow{ }^4F_{3/2}$ transition of Nd$^{3+}$ at approximately 879.7 $nm$ in the Nd$^{3+}$:YVO$_4$ crystal exhibits strong absorption of photons polarized parallel to the c-axis. The relevant energy level diagram of our experiment is shown in the inset in Fig. 1, where two Zeeman spin levels of ground state and an excited level constitute a $\Lambda$-like system. This transition exhibits a wideband inhomogeneous broadening ($\Gamma_{inh}\sim$ 2.1 GHz) and a narrow homogenous linewidth ($\Gamma_{h}\sim$ 63 kHz) at low temperature and a magnetic field \cite{Gisin08,Gisin10}. With the spectral-hole burning technique \cite{Gisin08,Gisin10}, the absorption profile can be user-programmed with a carefully designed pump beam \cite{polarization,lgi}. The lifetime of the obtained absorption profile is mainly determined by the population lifetime of the Zeeman spin levels under magnetic field, which is several $ms$ \cite{Gisin08,Gisin10}.

\begin{figure*}[tb]
\centering
\includegraphics[width=0.75\textwidth]{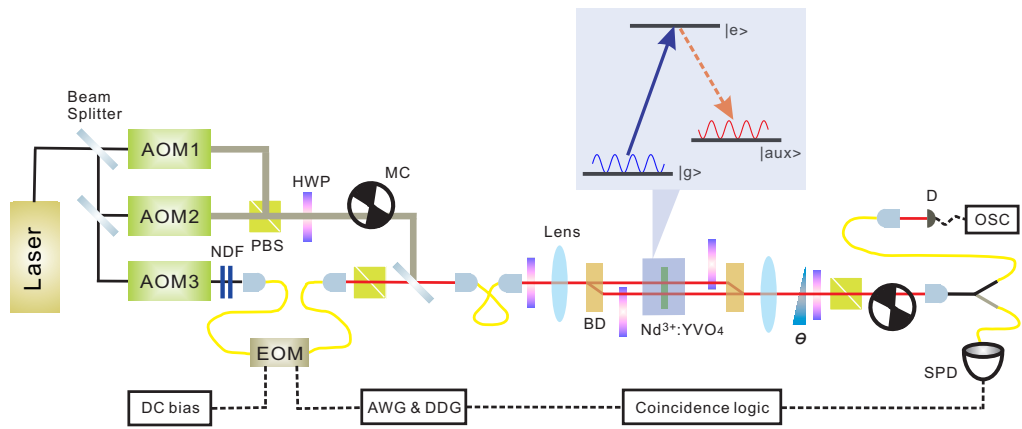}
\caption{\label{Fig:1} The experimental setup for observation of optical precursors in solids.
AOM1 (AOM2) produces $V$ ($H$) polarized pump light. The combination of AOM3 and EOM produces the weak probe pulses. The pump and probe light are collected with a SMF and then sent into the interferometer. After the sample, the two beams are again combined with another BD. The transmitted light's polarizations are then analyzed with a HWP and a PBS. To protect the SPD during the preparation procedure, two phase-locked mechanical choppers are placed in the pumping optical path and before the SPD. The inset shows the relevant energy level diagram of Nd$^{3+}$:YVO$_4$. One Zeeman spin level of ground state $|g\rangle$ and the excited state $|e\rangle$ are addressed with the pump and probe light. With continuous optical pumping, the absorption profile in frequency domain can be engineered through spontaneous emission and population trapping into the other Zeeman spin level $|aux\rangle$.}
\end{figure*}

Fig. 1 shows the experimental setup for investigating optical precursors using a polarization-based interferometer. The laser source is a continuous wave Ti:sapphire laser (M Squared, Solstis). The horizontal ($H$) and vertical ($V$) polarized pump light beams are independently generated with two 260 MHz acousto-optic modulators (AOM)  in double-pass configurations and combined with a polarization beam splitter (PBS). The probe light beam is generated by another 260 MHz AOM in double-pass configuration. The probe light is then sent to a high speed fiber-coupled electro-optical modulator (EOM, bandwidth: 15 GHz, EOspace). The EOM is driven by an 8 GS/s arbitrary wave generator (Tektronix AWG7082c) and a dc bias voltage. The probe light is decreased by neutral density filters (NDF). The pump light and probe light are combined with a beam-splitter and collected with a single-mode fiber (SMF) to ensure perfect mode overlap. The half-wave plate (HWP) placed after the SMF is used to prepare the probe light with a polarization of $H+V$. The first HWP placed in the pump path is used to correct the polarization of the pump light,
thus the pump light generated with AOM1 (AOM2) is simply $V$ ($H$) polarized before the lens.

The output from the SMF is focused onto the sample using the lens (focal length: 350 $mm$). The light is split into two beams depending on the optical polarization by a beam displacer (BD). The $H$ polarized beam (upper beam) is directly sent into the sample while the $V$ polarized beam (lower beam) is sent into the sample after the polarization is rotated to $H$ by a 45$^\circ$ HWP. Setting the polarization of the two beams to be identical before the beams are incident onto the sample ensures equal absorption for both beams. The sample is placed in a cryostat (Oxford Instruments, SpectromagPT) at a temperature of 1.5 K and with a superconducting magnetic field of 0.3 T in the horizontal direction. Beyond the cryostat, the polarization of the upper beam is rotated to $V$ by another 45$^\circ$ HWP. The two beams are combined again with another BD. A birefringent plate ($\theta$) is inserted in the probe path for fine-adjustment of the two polarization dependent delays. The HWP and PBS determine the polarization for detection. The transmitted light beams are collected with another SMF to ensure perfect mode matching of the two beams. The output is either detected using a 250 MHz photodetector or sent into a single-photon detector (SPD, PerkinElmer, SPCM AQRH-15). The output from the photodetector is recorded with a 5 GS/s oscilloscope (OSC, Tektronix DPO4104). The output of the SPD is analyzed with a coincidence counter (Ortec). A digital delay generator (DDG) provides the coincidence logic for the counter.

In the preparation phase, the pump light's frequency is swept over 100 MHz in every 100 $\mu$s and each frequency step has been assigned a specific amplitude to provide a user-determined structure. The preparation phase takes 9 ms. The probe phase starts after a waiting time of 1 ms. The complete pump and probe cycles are repeated at a frequency of 40 Hz.

\begin{figure}[tb]
\centering
\includegraphics[width=0.45\textwidth]{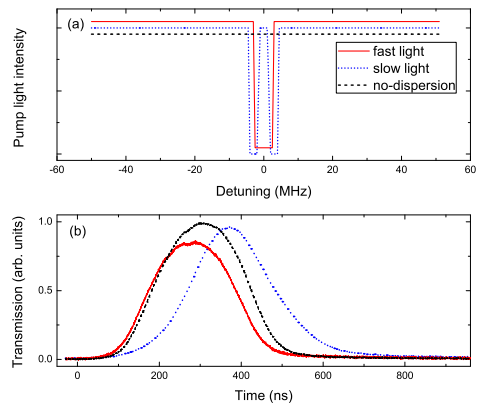}
\caption{\label{Fig:2} (a). Pump light profiles to obtain fast light (red solid), slow light (blue dotted) and no-dispersion conditions (black dashed). (b). Measured fast light and slow light propagation for a Gaussian pulse traversing through the programmable medium. The black dashed line, red solid line and blue dashed line represent the no-dispersion case, fast light case, and slow light case, respectively. All the traces are the average of 64 pulses. The fast light and slow light transmissions are intentionally magnified by a factor of approximately 2.3 for comparison purposes.}
\end{figure}

We first show that the crystal can be programmed into fast light and slow light regimes. According to the Kramers-Kronig relations, a spectral hole leads to slow light while an anti-hole in the absorption profile leads to fast light propagation \cite{Boyd03}. Slow light and fast light propagation in rare-earth-doped solids also have been achieved in previous experiments \cite{sl,sl2,fl}.
The pump light intensity as a function of the frequency scan is shown in Fig. 2(a). The black dashed line represents the no-dispersion case, where all ions in the 100 MHz range are pumped into other Zeeman spin levels; the sample shows a negligible and flat absorption for all of these frequency components. The blue dotted line represents pump light for the normal dispersion case, where a spectral hole is produced that is located at the central frequency. In this case, the probe light is in the slow light regime. The red solid line represents the pump light for the anomalous dispersion case, where an anti-hole is produced at the central frequency. In this case, the probe light will experience fast light propagation.
The pump light power ($<$ 200 $\mu$W) and the duty ratio in each cycle are carefully adjusted to take into account the power broadening effect. As a result, the sample exhibits a hole and anti-hole located at the central frequency in the absorption spectrum with a bandwidth $\Delta$ of approximately 6 MHz for the slow light and fast light cases, respectively. The peak absorption depth $d$ is approximately 0.9 (1.4) for the slow light (fast light) case.

The probe pulse is a Gaussian pulse with a full width at half maximum of approximately 250 ns. The power of the probe light is intentionally decreased to the order of 10 $\mu$W to avoid significant affects on the atomic absorption. Fig. 2(b) shows the corresponding measurements in the three different regimes. Compared with the no-dispersion case (black dashed line), the pulse peak experiences an advancement shift of approximately 23 ns for the fast light case. Considering the approximately $t_1\sim$20 ps propagation time for detuned light to traverse the 3-$mm$ length crystal, the 23-ns advancement indicates an effective group velocity of $\upsilon_{g}\simeq-1.3\times10^5$ $m$/s. The delayed pulse output in the slow light regime is represented by the blue dotted line. The group delay is determined to be approximately 72 ns using a simple fitting procedure, which corresponds to a group velocity of $\upsilon_{g}\simeq4.2\times10^4$ $m$/s.  The group delay for the on-resonance probe light can be estimated by $d/\Delta$ \cite{slow}, which roughly agrees with the directly measured group delay taking into account the small errors introduced by the background absorption ($d_0<0.4$). These results demonstrate that the crystal is highly controllable; one can switch the medium among the fast light, slow light, and the no-dispersion regimes by simply changing the corresponding pump light.

\begin{figure}[tb]
\centering
\includegraphics[width=0.45\textwidth]{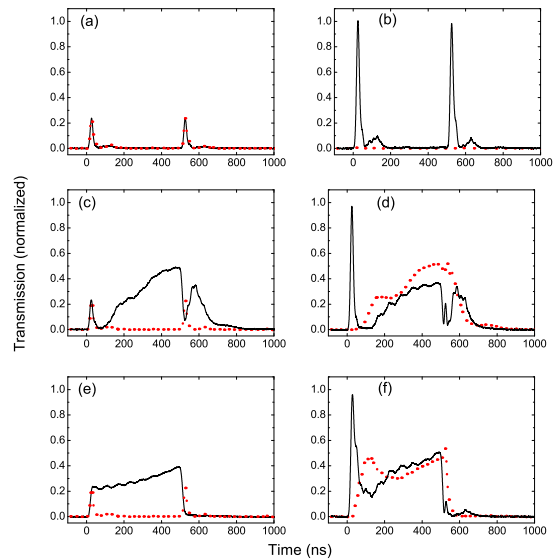}
\caption{\label{Fig:3} Interference between precursors that experienced different dispersive conditions. The black solid line in (a), (c) and (e) represents the $H-V$ polarized transmission of a probe pulse while the $V$- polarized pump light is blocked. The red dotted line represents the $H-V$ polarized transmission of the probe pulse while the $H$- polarized pump light is blocked. The $H$- polarized pump light is programmed to obtain fast light, slow light, and no-dispersion conditions in (a), (c) and (e), respectively. (b), (d), and (f) show the corresponding interference results with the pump light described in (a), (c) and (e), respectively. The red dotted (black solid) line represents the $H-V$ ($H+V$) component of transmitted pulse. All the traces are the average of 64 pulses. }
\end{figure}

In the following experiments, the two parts of the crystals in the two arms of the interferometer are independently pumped by the $H$- and $V$-polarized pump light. The probe pulse generated by the EOM is changed to a square pulse with temporal width of 500 ns and a rise time $<$ 0.5 ns. We first program the $H$- and $V$-polarized pump light to obtain the same anomalous dispersions for the two beams. The corresponding transmissions are shown in Fig. 3(a). The polarization of detection is chosen as $H-V$ by tilting the birefringent plate and adjusting the angle of the last HWP. The black solid line shows the transmission recorded in the oscilloscope, while the $V$- polarized pump beam is blocked. Because of the high optical depth of the sample ($\sim12$) and the wideband inhomogeneous broadening, without the $V$-polarized pump light, the $V$-polarized probe pulse is almost completely absorbed by the sample. So the recorded transmissions are $H$-polarized pulses. The $V$-polarized transmission represented by the red dotted line, is obtained by blocking only the $H$- polarized pump light. The pump power is carefully adjusted to achieve the balanced transmissions for the two polarization states, which is necessary to explore the maximum visibility of transmitted pulses. The main fields are almost totally absorbed because of the large on-resonance absorption ($d\sim$ 4). Because the fast rising/falling edges contain wideband frequency components which are beyond the atomic absorption band, the transmissions show sharp peaks at $t=0$ and $t=500$ ns. Similar results have been observed previously in cold atoms \cite{Du09}. The measured transmissions of $H-V$-polarized light when both the $H$- and $V$-polarized pump light are incident on the sample are represented by the red dotted line in Fig. 3(b). Nearly perfect destructive interference between the two beams is observed. By rotating the HWP, one can choose the polarization of $H+V$ for detection. The corresponding transmissions are represented by the black solid line in Fig. 3(b), which indicates nearly perfect constructive interference.

To determine whether the polarization of the precursors is affected by strong dispersions in the medium of propagation, we program the $H$-polarized pump light to obtain slow light conditions for the upper path of the interferometer. The $V$-polarized pump light remains unchanged. The corresponding transmission of the probe pulse with $V$- ($H$-)polarized pump light blocked is shown in Fig. 3(c) by the black solid (red dotted) line. For the slow light case, the precursors still arrive at the same time as the fast light case, while the main field arrives approximately 100 ns later.  The interference results are shown in Fig. 3(d), where the fast rising wavefronts remains a minimum for the $H-V$ polarized detections and approaches a maximum for the $H+V$ polarized detections.

We further show the transmissions for the $H$-polarized probe pulses in Fig. 3(e) when the crystal in the upper path is programmed for the no-dispersion condition. For perfect optical pumping that the medium shows no absorption for the input pulse, the transmission should be a square pulse. However, limited by the short Zeeman lifetime and small branching ratio, there is always a small background absorption ($d_0<0.4$) because of inefficient optical pumping. The small absorption is easily saturated by the classical probe pulses. So the transmission deviates a little from a square pulse with small saturation behavior. The small oscillating behavior of the pulse is caused by the non-linear response of the electro-optical modulating system. The $V$-polarized pump light remains unchanged to achieve the fast light regime. The interference results shown in Fig. 3(f) again confirm that the polarization of the rising edges remains unchanged.

In the Supplemental Information, we have presented a quantitative analysis of polarization states of wavefronts using single-photon level probe pulses. The results show that the wavefronts' polarization remains unchanged with a high visibility of 99.8\% for all three dispersion conditions. These results demonstrated that the polarization states of wavefronts show little dependence on the media dispersions. Our scheme can also be performed in other systems where no background medium is present, such as cold atoms and room temperature atom vapors. In that case, the vacuum can be taken as the reference medium where the speed of light is well defined. The speed of optical precursors then can be compared with an absolute bound.

We demonstrated the ability to obtain the fast light and slow light propagation in user-programmable rare-earth-doped solids. While the main fields are greatly delayed or accelerated by the mediums, the wavefronts travel at a constant speed and exhibit little dependence on the dispersion conditions. The unaltered polarization-encoded information carried by the precursors suggests that optical precursors are more robust against decoherence caused by the environment compared with the main fields, and therefore should find applications in both quantum and classical communication networks.

This work was supported by the CAS, the National Basic Research Program (2011CB921200), National Natural Science Foundation of China (Grant No. 60921091 and No. 11274289) and the Fundamental Research Funds for the Central Universities (wk2030380004 and wk2470000011).

\section*{Supplemental Information}
\subsection*{Analysis of wavefronts' polarization using a single photon detector (SPD).}

We notice that the minimum output of the dark port detection ($H-V$ polarization) is beyond the resolution of our classical detectors. To maximally explore the high visibility of the interferometer, we decrease the probe pulse to the single-photon level and the transmitted light is detected using the SPD (PerkinElmer, SPCM AQRH-15). The single-photon experiments also have the advantage of negligible affections on the atomic absorption spectrum. Because only the wavefront parts of the transmission is of interest, we place the coincidence window approximately in the range [-1 ns, 2.5 ns].

We analyzed the wavefront polarization as obtained in Fig. 3(b) using SPD. 500 single-photon levlel pulses are generated at a frequency of 800 KHz in each probe cycle. The coincidence counts as a function of the half-wave-plate (HWP) angle is shown in Fig. 4. We observe a maximum count rate of approximately 2200 per second for $H+V$ polarization and a minimum count rate of 2 per second for $H-V$ polarization. The results are almost the same for all the 3 dispersion conditions of upper path. So we only show the fast light condition for clarity.

\begin{figure}[tb]
\centering
\includegraphics[width=0.5\textwidth]{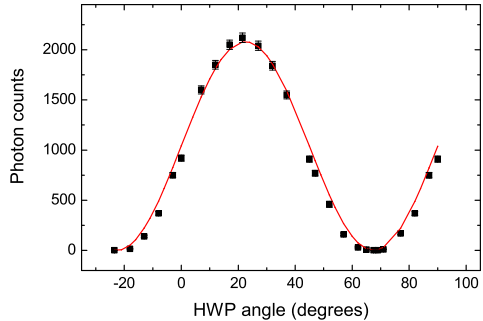}
\caption{\label{Fig:s1} The analysis of the polarization states of the transmitted wavefronts using SPD. Both the $H$- and $V$-polarized pump light are programmed to obtain fast light conditions. The photon counts exhibit a maximum of 2200 for $H+V$ polarization and a minimum of 2 for $H-V$ polarization.}
\end{figure}

Without any further adjustment to any optical elements, we then switch the $H$-polarized pump light to obtain slow light dispersion conditions. The photon count rate of the dark port remains low at 2 counts per second. The $H$-polarized pump light is then switched to obtain no-dispersion condition, and the dark port photon count rate remains a minimum. For all three cases, we found that the dark port photon counts remain a minimum at $\leq$ 2 counts per second, corresponding to a visibility of 99.8\%. We note that the interference visibility is mainly limited by the vibrations of sample under low temperature. The beam-displacer-based interferometer exhibits an extinction ratio of P$_{max}$/P$_{min}$ of $>$ 20000:1 if the cryostat is not running. When the sample is at low temperature, the pulse tube refrigerator introduces the largest amount of noise to the interferometer. Carefully tilting the sample so that the light is incident perpendicularly to the sample surface can minimize the influence caused by the sample vibrations. The extinction ratio of the interferometer with the sample at the base temperature is P$_{max}$/P$_{min}$ of $>$ 8000:1, for continuous light.


\end{document}